\documentclass[12pt]{article}
 \newcommand{\be}{\begin{eqnarray}}
 \newcommand{\ee}{\end{eqnarray}}
 \newcommand{\nee}{\nonumber\end{eqnarray}}
 \newcommand{\nn}{\nonumber}
\usepackage{amsthm,amssymb,amsfonts}

\begin{document}

\begin{titlepage}
\begin{flushright}
hep-ph/0007303\\ CERN-TH/2000-177\\ July 2000
\end{flushright}
\vfill
\begin{center}
{\Large\bf A Strategy for the Analysis of \\ Semi-Inclusive Deep
Inelastic Scattering}
\\
\vspace{2cm}
{\large\bf Ekaterina Christova}
\footnote{Permanent address:
\rm Institute of Nuclear Research and Nuclear Energy,
Sofia,\\
 e-mail: echristo@inrne.bas.bg}\\ {\em Theory Division, CERN, CH
1211 Geneva 23}

\vspace{1cm} {\large\bf Elliot Leader} \footnote {Permanent
address: \rm High Energy Physics Group, Imperial College,
London,\\ e-mail: e.leader@ic.ac.uk}\\ {\em Vrije Universiteit,
Amsterdam}
\end{center}
\vfill

\begin{abstract}

We present a strategy for the systematic extraction of a vast
amount
 of detailed information on polarized parton densities and
fragmentation functions from semi-inclusive deep inelastic
scattering $l+N\to l+h+X$, in both LO and NLO QCD. A method is
suggested for estimating the errors involved in the much simpler,
and therefore much more attractive, LO analysis. The approach is
based upon a novel interplay with data from inclusive DIS and from
$e^+e^-\to hX$, and leads to a much simplified form of the NLO
expressions. No assumptions are made about the equality of
 any parton densities and the only symmetries utilised are charge
 conjugation invariance and isotopic spin invariance of strong
 interactions.

\end{abstract}
\end{titlepage}
\newpage
\setcounter{page}{1}


\section{Introduction}


There are two major problems in the QCD analysis of polarized
{\it inclusive} deep inelastic scattering (DIS): i) the absence
of neutrino data makes it impossible, in principle, to determine
the non-strange polarized sea-quark densities $\Delta\bar
u(x,Q^2)$ and $\Delta\bar d(x,Q^2)$, ii) the separate
determination of the polarized strange quark density $\Delta
s(x,Q^2)$ and the polarized gluon density $\Delta G(x,Q^2)$
relies heavily on the QCD evolution in $Q^2$ and use of the
flavour $SU(3)_F$-invariance relation
\be
\int_0^1\,dx \left[\Delta u+\Delta\bar u +\Delta d +\Delta\bar d
-2\Delta s -2\Delta \bar s\right] = 3F-D. \ee
 The absence of a long lever arm in $Q^2$, in the polarized case,
and doubt concerning the reliability of $SU(3)_F$-invariance for
hyperon $\beta$-decay
 means that $\Delta s$ and $\Delta G$ are still rather poorly
known \cite{Altarelli}, despite the dramatic improvement in the
quality of the data in the past few years.

The direct resolution of these problems must await a series of
new machine development projects, based on very high intensity
neutrino beams,
  which are most unlikely to come into operation before the year
2015.

In the meantime there is currently a major experimental effort at
CERN \cite{SMC}, HERA \cite{HERMES} and Jefferson Lab. to study
semi-inclusive polarized DIS reactions
\be
 \overrightarrow{e} + \overrightarrow{N}\to e+ h+ X,\label{theprocess}
\ee
involving the detection of the produced hadron $h$.

The theoretical structure for the analysis of such reactions, in
both leading order QCD (LO) and next-to-leading order (NLO),
exists \cite{deFlorian}. However we are critical of the type of
LO analysis carried out thus far by the experimental groups.

The analysis of semi-inclusive DIS \cite{SMC,HERMES} involves both
parton densities and fragmentation functions. In the LO treatments
referred to above, the fragmentation functions (FF's) are treated
as known quantities from inclusive $e^+e^-\to hadrons$, and are
used in constructing an auxiliary quantity called "purity".
However, it is well known that in $e^+e^-\to hadrons$, both in LO
and NLO, first, only the combinations $D_q^h + D_{\bar q}^h$ can
be determined while for the analyses of semi-inclusive DIS both
$D_q^h$ and $D_{\bar q}^h$ are needed separately, and second, the
existing analysis are rather ambiguous: a detailed study in
\cite{Binnewies} makes a 31 parameter fit to the data, and no
errors are quoted, and, in a more recent study \cite{Kretzer},
 the FF's differ significantly from those in
\cite{Binnewies}, by 40\% or more in some regions of $z$. Under
these circumstances it is unreasonable to pretend to have an
absolute knowledge of the fragmentation functions.

 Two NLO analyses based upon a
global analysis of the inclusive and semi-inclusive data have been
attempted \cite{NLOanalysis}, \cite{deFloriannew}. In the more
recent analysis
 \cite{deFloriannew} the authors relax the equality
of $\Delta \bar u(x)$ and $\Delta \bar d(x)$ and find a preference
for a positive $\Delta \bar u(x)$,
 but effectively no constraint on the sign of $\Delta \bar d(x)$.
However, in both these analyses it is again assumed that the FF's
are known exactly: those of \cite{Binnewies} being used in
 \cite{NLOanalysis}, and those of  \cite{Kretzer} in \cite{deFloriannew}.

In addition, in the above mentioned analysis, some simplifying
assumptions are made about relations between various polarized
parton densities. In the following, except where expressly
indicated, we make no assumptions at all concerning the polarized
or unpolarized parton densities.
 Indeed there are persuasive arguments from the large-$N_c$ limit
of QCD that a significant difference should exist between $\Delta
\bar u(x)$ and $\Delta \bar d(x)$ \cite{Diakonov} with $\vert
\Delta \bar u - \Delta \bar d\vert > \vert\bar u-\bar d\vert$,
and it has been argued that such a situation is compatible with
all present day data~\cite{Dressler}. Further, bearing in mind
recent arguments~\cite{a} that
 $s(x)\not= \bar s(x)$ and $\Delta s(x)\not= \Delta \bar s(x)$,
we even refrain from the very common assumption of the {\it
equality of these densities}.

 However, from experience gained in
the analysis of inclusive polarized DIS, it appears that the
parameter space is sufficiently complicated to be able to produce
biases in the $\chi^2$ analysis, which can lead to unphysical
results. We thus believe it to be dangerous in either LO or NLO
QCD, to put together all inclusive and semi-inclusive data in one
global analysis. Rather, what is required, is a
 working strategy, making optimal use of selected parts of the
data.

The aim of this paper is precisely to provide a strategy, in both
LO and NLO QCD, for the analysis of the semi-inclusive data. We
proceed as follows:  Information about the FF's is obtained from
{\it unpolarized} semi-inclusive data. Then this information is
used to determine the polarized parton densities from {\it
polarized} semi-inclusive DIS. Appropriate use of the information
from DIS and $e^+e^- \to hadrons$ is made as well. We suggest, for
example, that one should use as {\it input} not just a knowledge
of the unpolarized parton densities and their errors, but rather
the polarized isotopic combination
\be
\Delta q_3(x,Q^2) = (\Delta u+\Delta \bar u) -(\Delta
d+\Delta\bar d)\label{Deltaq3}
\ee
 which is by now very well determined from polarized inclusive
data, and which is free from any influence of $\Delta s$ and
$\Delta G$. Inclusive $e^+e^- \to hadrons$ and DIS are used also
to considerably simplify the NLO expressions for the cross
sections of (\ref{theprocess}).

 Further, in stead of dealing with
each parton density and FF separately, as is often done we work
with their singlet and non-singlet combinations.  As both the
parton densities and the FF's enter the semi-inclusive cross
sections, we single out such observables that are singlet
(non-singlet) combinations on both the parton densities and the
FF's.

This leads to the fact  that we often consider linear combinations
of experimentally measured
 quantities, and it may be objected that thereby we
 are dealing with experimental observables with possibly large
errors. It has to be understood that that is a reflection of the
true situation and not an artifact of our approach. To take an
absolutely trivial example, suppose $E_1$ and $E_2$ are two
experimentally measured functions used in an
 attempt to determine the theoretical functions $T_1$ and $T_2$,
where $E_1 = T_1 + T_2$ and $E_2 = T_1 - T_2$. Now, if it happens
that $E_1 \sim E_2$ and if we write $T_{1,2} = (1/2) (E_1\pm E_2)$
then $T_2$ will be very poorly determined. This is unavoidable. It
does not help to do a best fit to $E_1$ and $E_2$ with some
parametrisations of $T_1$ and $T_2$. {\it Inherently} the result
for $T_2$ will have a large relative error.

Thus we believe that any relatively large errors occurring in our
manipulation of the experimental quantities reflects a genuine and
unavoidable imprecision
 in the determination of certain theoretical quantities.

In order to minimize systematic errors experimentalists prefer to
consider asymmetries or ratios of cross sections where the
detection
 efficiencies should roughly cancel out, e.g. ratios of polarized
to unpolarized DIS or ratios of polarized to unpolarized
semi-inclusive for a given detected hadron.
 We appreciate that this is a fact of experimental life. However
we wish to stress that a large amount of information is lost in
restricting oneself to only
 these ratios. It is vitally important to gain control of the
systematic errors in detection efficiencies, and although we try
as far as possible to deal with the favoured kind of ratios
 we will be forced also to consider other types of cross-section ratios.

Throughout the rest of this paper we assume that a kinematic
separation is possible between hadrons produced in the current
fragmentation region and those produced from the target remnants.
We consider only the current fragmentation region so that our
formulae apply only to this region and fracture functions
\cite{32} play no role in our discussion.

It has to be stressed that there is a huge difference in
complexity between the LO and NLO treatments. Thus it makes sense
to utilize the LO approach,
 provided appropriate checks, (which we suggest) are carried
out.

Our analysis proceeds in a step-by-step fashion. Firstly we
describe a generic
 test for the reliability of a LO treatment. Assuming this to be
successful we present the analysis in LO. In LO the information on
the polarized valence quark densities $\Delta u_V$ and $\Delta
d_V$ and on the breaking of SU(2) invariance of the polarized sea
is obtained without any knowledge of the FF's. However, it should
be clear that any  information on the sea quark densities in LO
cannot be reliable,  since they are expected to be small and thus
comparable to the NLO corrections. We consider also the
experimentally
 difficult case of $\phi$ production, which seems to be the best
 way to get an accurate determination of $\Delta s+\Delta\bar s$
 in LO. We discuss also how, in principle, one can test
  $s(x)= \bar s(x)$ and/or $\Delta s(x)=\Delta\bar s(x)$.

In the NLO treatment we show in a new way how information on
$e^+e^- \to h+X$ and
 inclusive DIS can be incorporated directly so as to considerably
 simplify the
NLO expressions for semi-inclusive cross sections. This key result
is presented in eqs. (\ref{sigmaNLOnew}) and
(\ref{deltasigmaNLOnew}).  Next we discuss the fragmentation
combination $D_u^{\pi^+} - D_u^{\pi^-}$ and use it to evaluate
 $\Delta u_V$ and $\Delta d_V$ in NLO. Then we obtain expressions
for $D_u^{\pi^+} + D_u^{\pi^-}$, for $D_G^{\pi^+}$ and for
$D_s^{\pi^+} + D_s^{\pi^-}$. With these we are able to obtain
($\Delta\bar u - \Delta\bar d$) in NLO. Finally we have a set of
2 equations
 involving 3 unknown functions, ($\Delta u + \Delta\bar u +
\Delta d + \Delta\bar d$), ($\Delta s+\Delta\bar s$) and $\Delta
G$. An {\it accurate} determination of all three functions would
require data over a presently impossibly wide range of $Q^2$. We
thus suggest using here for $\Delta G$ its determination from
charm
 production \cite{charm}. It should then finally be possible to get an
 accurate assessment of $\Delta s+\Delta\bar s$ in NLO. Lastly we
 consider the evaluation of $ s(x)-\bar s(x)$ and $\Delta
 s(x)-\Delta\bar s(x)$.


\section{A strategy for semi-inclusive DIS}


In NLO QCD, the expressions for semi-inclusive cross sections
involve {\it convolutions} of parton densities and fragmentation
functions with (known) Wilson coefficients . Our lack of knowledge
of the errors on the FF's will make it difficult to assess the
accuracy of the parton densities which we are trying to determine.
In the LO QCD approximation, on the other hand there are no
convolutions, but simple products only {\it (independent
fragmentation)}, and it becomes possible to construct measurable
combinations of cross sections in which the FF's completely cancel
out \cite{Frankfurt,we}. However it is not clear how reliable the
LO is.

We believe it is quite safe when determining the large $\Delta u$
and $\Delta d$ densities, but could be quite misleading for
$\Delta\bar u$, $\Delta\bar d$ and $\Delta s$. In any event it is
absolutely essential to {\it test} independent fragmentation in
order to have a feeling for the errors on parton densities
obtained via the LO formalism.

In LO, the structure of the expressions is generally of the form
\be
parton\,\, density\,\, \Delta q(x,Q^2) = experimental\,\,
 observable\,\,E(x,z,Q^2)\label{parton}
\ee
or
\be
fragmentation\,\, function\,\,D(z,Q^2) = experimental\,\,
 observable\,\, E(x,z,Q^2).\label{fragm}
\ee
In both cases the characteristic feature of the LO treatment is
that the RH sides, which can in principle depend on $(x,z,Q^2)$,
should only depend on two of these, either $(x,Q^2)$ or $(z,Q^2)$
respectively, so that there is an independence of the third
variable, which we shall call the {\it passive} variable.

Every expression of the form (\ref{parton}) or (\ref{fragm})
should be tested for dependence on the passive variable. If a
significant dependence is found it does not mean that the LO
analysis must be abandoned, but it suggests that the variation
with the passive variable be used as an estimate of the {\it
theoretical} errors, $\delta_{TH} [\Delta q(x,Q^2)]$,
 $\delta_{TH} [D(z,Q^2)]$ on the sought for quantities.

In Section 7 we discuss a strategy for the analysis in NLO.


\section{Parton densities in LO QCD}


It is useful to introduce the following notation for
semi-inclusive processes:
\be
\tilde \sigma^h \equiv \frac{x(P+l)^2}{4\pi
\alpha^2}\left (\frac{2y^2}{1+(1-y)^2}\right )
\frac{d^3\sigma^h}{dx\,dy\,dz}\label{sigmah}
\ee
and
\be
\Delta \tilde \sigma^h \equiv \frac{x(P+l)^2}{4\pi
\alpha^2}\left (\frac{y}{2-y}\right ) \left
[\frac{d^3\sigma_{++}^h}{dx\,dy\,dz} -
\frac{d^3\sigma_{+-}^h}{dx\,dy\,dz}\right ]\label{Deltasigma}
\ee
where $P^\mu$ and $l^\mu$ are the nucleon and lepton four
momenta, and $\sigma_{\lambda\mu}$ refer to a lepton of helicity
$\lambda$ and a nucleon of helicity $\mu$. The variables $x$,
$y$, $z$ are the usual DIS kinematic variables.

Then in LO the cross sections for the semi-inclusive production
of a hadron $h$ have the simple $y$-independent form
\be
\Delta \tilde \sigma^h (x,z,Q^2) &=& \sum_{q,\bar q} e^2_q\,
\Delta q_i(x,Q^2)\,
D_i^h(z,Q^2) \\
\tilde \sigma^h (x,z,Q^2) &=&  \sum_{q,\bar q} e^2_q \,q_i(x,Q^2)\,
D_i^h(z,Q^2),\label{sigmaLO}
\ee
where the sum is over quarks and aniquarks, and where $D_i^h$ is
the fragmentation function for quark or antiquark $i$ to produce
$h$.

We consider sum and difference cross sections for producing $h$
and its charge conjugate $\bar h $ on both protons and neutrons,
and define
\be
\Delta A_{p,n}^{h\pm \bar h }(x,z,Q^2) &=&
\frac {\Delta\tilde\sigma_{p,n}^{h}
\pm\Delta\tilde\sigma_{p,n}^{\bar h }}
{\tilde\sigma_{p,n}^{h}
\pm \tilde\sigma_{p,n}^{\bar h }}
\equiv
\frac {\Delta\tilde\sigma_{p,n}^{h\pm \bar h }}
{\tilde\sigma_{p,n}^{h\pm \bar h }},\label{1}\\
\Delta A_{p\pm n}^{h\pm \bar h }(x,z,Q^2) &=&
\frac {\Delta\tilde\sigma_p^{h\pm \bar h } \pm
\Delta\tilde\sigma_n^{h\pm \bar h }}
{\tilde\sigma_p^{h\pm \bar h } \pm\tilde\sigma_n^{h\pm \bar h
}}.\label{2}
\ee

For inclusive unpolarized and polarized DIS cross sections we use
the notation:
\be
\tilde \sigma^{DIS}
\equiv \frac{x(P+l)^2}{4\pi
\alpha^2}\left (\frac{2y^2}{1+(1-y)^2}\right )
\frac{d^2\sigma^{DIS}}{dx\,dy}\label{sigmaDIS}
\ee
and
\be
\Delta \tilde \sigma^{DIS} \equiv \frac{x(P+l)^2}{4\pi
\alpha^2}\left (\frac{y}{2-y}\right ) \left
[\frac{d^3\sigma_{++}^{DIS}}{dx\,dy} -
\frac{d^3\sigma_{+-}^{DIS}}{dx\,dy}\right ].
 \ee
 Then in LO we
have the $y$-independent expressions:
\be
\tilde\sigma^{DIS} (x,Q^2)& =& 2F_1^N (x,Q^2)\vert_{LO}=
\sum_{q,\bar q} e^2_q \,q_i(x,Q^2)\,\\
 \Delta \tilde\sigma^{DIS}
(x,Q^2) &=& 2g_1^N (x,Q^2)\vert_{LO}= \sum_{q,\bar q} e^2_q
\,\Delta q_i(x,Q^2)\,.
 \ee

In addition to (\ref{1}) and (\ref{2}) we consider the ratios
 of sum and difference hadron yields for the unpolarized
semi-inclusive and inclusive processes:
\be
 R_{p,n}^{h\pm \bar h }(x,z,Q^2) =
\frac {\tilde\sigma_{p,n}^{h}\pm  \tilde\sigma_{p,n}^{\bar h }}
{\tilde\sigma^{DIS}_{p,n }},\qquad R_{p\pm n}^{h\pm \bar h
}(x,z,Q^2) =
\frac {\tilde\sigma_p^{h\pm \bar h } \pm
\tilde\sigma_n^{h\pm \bar h }}
{\tilde\sigma_p^{DIS } \pm\tilde\sigma_n^{DIS }}.\label{R}
\ee
(It is equally good to use a sum over any set of hadrons $h$ and
their charge conjugate $\bar h $.)


\subsection{Testing LO QCD}


Using only charge conjugation invariance it is easy to show that
\cite{we}
\be
\Delta A_{p-n}^{h + \bar h }(x,z,Q^2) =
\frac{\Delta\tilde\sigma^{DIS}_p - \Delta\tilde\sigma^{DIS}_n}
{\tilde\sigma^{DIS}_p - \tilde\sigma^{DIS}_n}(x,Q^2)= \frac{(g_1^p
- g_1^n)\vert_{LO}}{(F_1^p-F_1^n)\vert_{LO}} (x,Q^2) .\label{R++}
\ee This is a key relation for testing the reliability of the LO.
The RHS is completely known from inclusive DIS and is independent
of $z$.
 The LHS, in principle, depends also upon $z$ and upon the hadron
$h$. Only in LO (or in the simple parton model) should it be
independent of $z$ and of $h$. It is thus crucial to test this
feature.

To help with statistics it is also possible to formulate an
integrated version of (\ref{R++}). This is given in \cite{we}.
For the rest of this section we assume that the test (\ref{R++})
has been successful and proceed with the analysis in LO.


\subsection{The valence quark densities in LO}


The polarized valence quark densities can be obtained from
$\pi^\pm$ production, assuming only isospin invariance:
\be
\Delta u_V &=& \frac{1}{15}\left\{4(4u_V - d_V)\Delta A_{p}^{\pi^+-
\pi^-}
+ (4d_V -u_V)\Delta A_{n}^{\pi^+-\pi^-}\right\}\nn\\
\Delta d_V &=&\frac{1}{15}\left\{4(4d_V - u_V)\Delta A_{n}^{\pi^+-
\pi^-}
+ (4u_V - d_V)\Delta A_{p}^{\pi^+-\pi^-}\right\}.\label{A}
\ee

For the case of $K^{\pm}$ or $\Lambda,\bar\Lambda$ production,
 if one makes also the conventional assumption that $s=\bar s$
and $\Delta s = \Delta\bar s$ one has in addition:
\be
\Delta u_V &=& \frac{1}{2}\left\{(u_V + d_V)\Delta A_{p+n}^{K^+ - K^- }
+ (u_V -d_V)\Delta A_{p-n}^{K^+ - K^- }\right\}\nn \\
\Delta d_V &=&\frac{1}{2}\left\{(u_V + d_V)\Delta A_{p+n}^{K^+ - K^-}
- (u_V - d_V)\Delta A_{p-n}^{K^+ - K^- }\right\}.\label{B}
\ee
Note that we have not assumed $D_d^{K^+}= D_d^{K^-}$, although
that
 is suggested by the absence of a $d$ quark in the leading Fock
state of $K^\pm$. Indeed the above equality can be tested (see
Section 5.2).

For isoscalar hadrons like $\Lambda ,\bar\Lambda$ again assuming
$s=\bar s$ and $\Delta s = \Delta\bar s$, one finds
\be
\Delta u_V &=& \frac{1}{15}\left\{4(4u_V +d_V)
\Delta A_{p}^{\Lambda -\bar\Lambda}
- (4d_V + u_V)\Delta A_{n}^{\Lambda -\bar\Lambda}\right\}\nn\\
\Delta d_V &=&\frac{1}{15}\left\{4(4d_V + u_V)
\Delta A_{n}^{\Lambda -\bar\Lambda}
- (4u_V + d_V)\Delta A_{p}^{\Lambda -\bar\Lambda}\right\}.
\label{C}
\ee
We shall comment in Section 5.5 on the situation if one does not
assume $s=\bar s$ and $\Delta s = \Delta\bar s$, where we suggest
a method for
 estimating if the failure of these equalities is serious or not.
In any event, the safe way to obtain $\Delta u_V$ and $\Delta
d_V$ is via $\pi^\pm$ production, eq. (\ref{A}).

Once again the reliability of these LO equations can be tested by
checking that the RH sides of (\ref{A}), (\ref{B}) and (\ref{C})
do not depend on $z$. If it is found that for a given $x$-bin the
RH sides vary with $z$ by some amount $\delta_{TH}[\Delta u_V]$,
$\delta_{TH}[\Delta d_V]$, then these could be regarded as an
estimate of the theoretical error at this $x$ value.


\section{Use of $\Delta q_3(x,Q^2)$ from polarized inclusive DIS}


At NLO the spin dependent structure functions for protons and
neutrons are given by:
 \be g_1^p (x,Q^2) &=&\frac{1}{2}\sum_{q=u,d,s}e^2_q
 \left[(\Delta q+\Delta\bar q) \otimes
 \left(1+\frac{\alpha_s(Q^2)}{2\pi}\delta C_q\right)+
 \right.\nn\\ && \hspace{5.5cm}+
 \left.\frac{\alpha_s(Q^2)}{2\pi}\Delta G\otimes
  \delta C_G \right]\label{g1p}
   \ee
    involving a convolution of polarized parton densities with
known Wilson
 coefficients. For the neutron,
    $g_1^n$ is obtained by the replacement \be \Delta u
\Longleftrightarrow\Delta d \ee in $g_1^p$.

Now it is clear that one can  obtain information only on the
combinations:
\be
\Delta u+\Delta\bar u, \quad \Delta d+\Delta\bar d,\quad \Delta
s+\Delta\bar s, \quad \Delta G.\label{deltaq}
 \ee
 and that it is
impossible to obtain separate information on the valence and
non-strange sea quark polarizations from inclusive, neutral
current, polarized DIS. \footnote{However, sometimes it is
convenient to parametrize separately the valence and sea quarks
and to assume, for example, $\Delta \bar u = \Delta \bar d
=\lambda \Delta s$, where $\lambda$ is a free parameter. It should
be obvious then, that any claim that the $\chi^2$ analysis favors
some particular value of $\lambda$ must be fictitious and a
consequence of some hidden bias in the minimization procedure. Yet
such claims have been made.}

In our semi-inclusive analysis we shall use as a known qauntity
 {\it only} the non-singlet combination of the polarized parton densities
$\Delta q_3$, eq.(\ref{Deltaq3}).  Now that there is such an
improvement in the quality of the neutron data we believe this
quantity is very well constrained directly by the inclusive data.
For one has, from (\ref{g1p})
\be
g_1^p(x,Q^2) -g_1^n(x,Q^2) =\frac{1}{6} \Delta q_3\otimes \left(1+
\frac{\alpha_s(Q^2)}{2\pi}\delta C_q\right) \ee and $\Delta
q_3(x,Q^2)$ is determined without any influence from the less well
known quantities $\Delta s$ and $\Delta G$, either in LO or in
NLO.  Of course if the semi-inclusive analysis is done in LO one
must use $\Delta q_3(x,Q^2)\vert_{LO}$. Note that we do {\it not}
use information on $(\Delta u +\Delta\bar u)$ or $(\Delta d
+\Delta\bar d)$ from inclusive DIS, since these are subject to the
less known  strange quark and gluon effects.


\subsection{SU(2) symmetry of the sea quark densities in LO}


One has
\be
[\Delta \bar u(x,Q^2)-\Delta \bar d(x,Q^2)]_{LO} =
\frac{1}{2}\left[\Delta q_3(x,Q^2) + \Delta  d_V(x,Q^2)-
\Delta  u_V(x,Q^2)\right]_{LO}.\label{sea}
\ee
   Eq. (\ref{sea}) determines the SU(2) symmetry breaking of the
 polarized sea
  without requiring any knowledge of the unknown $\Delta \bar q$
and $\Delta G$.
 A possible test for SU(2) breaking for the polarized sea
 densities that does not require any knowledge of the polarized
 densities was given in \cite{we}.

In order to determine the polarized sea quark densities
  separately we need one more relation, namely the value of
\be
 \Delta q_+(x,Q^2)\equiv \Delta u +\Delta\bar u +\Delta
d+\Delta\bar d.\label{deltaq+}
\ee
For then
\be
(\Delta\bar u +\Delta\bar d)_{LO} =
\frac{1}{2} (\Delta q_+ -\Delta u_V -\Delta d_V)_{LO}\label{sea2}
\ee
which combined with (\ref{sea}) yields $\Delta\bar u$ and
 $\Delta\bar d$ separately.

Although each term on the RHS of (\ref{sea}) and (\ref{sea2})
should be well determined in LO, the corresponding linear
combinations are expected
  to be small and may thus be very sensitive to NLO corrections.
An indication of the sensitivity may be inferred from the fact
that in inclusive polarized DIS, $\Delta s(x,Q^2)$ changes by
roughly a factor of 2 in going from LO to NLO or when one changes
factorisation schemes from $\overline{MS}$ to $AB$ or $JET$.

Note that determining, say, $\Delta\bar u$ via $\Delta\bar u =
{1\over 2} [(\Delta u+\Delta\bar u)-\Delta u_V]$ is
 unreliable since determination of ($\Delta u+\Delta\bar u$) from
inclusive DIS requires a knowledge of $\Delta s$.

In order to determine $\Delta q_+(x,Q^2)$ it will first be
necessary to extract some information on the FF's.

\subsection{Fragmentation functions in LO}


{\bf 1.} From measurements of the ratios $R_{p-n}^{h+\bar h}$ of
the semi-inclusive to inclusive DIS cross sections on protons and
neutrons for any given hadron $h$,
   it is feasible in LO to learn a great deal about the
FF's $D_q^h + D_q^{\bar h} \equiv D_q^{h+\bar h}$. This
combination is measured also  in $e^+e^- \to hadrons$.

Analogous to the polarized case, we define
\be
q_3(x,Q^2)& =& u(x,Q^2) + \bar u(x,Q^2) - d(x,Q^2) - \bar
d(x,Q^2)\\
  q_+(x,Q^2) &=& u(x,Q^2) +\bar u(x,Q^2) + d(x,Q^2) +\bar
d(x,Q^2).\label{q+}
\ee
which are well determined from inclusive DIS data and which thus
can be taken as known quantities in the semi-inclusive analysis.
\begin{itemize}
\item
 Using data on unpolarized semi-inclusive DIS we have, in LO,
\be
R_{p-n}^{h+\bar h} = \frac{2}{3}
\left[ 4D_u^{h + \bar h }(z,Q^2) -D_d^{h +\bar h }(z,Q^2)\right]
\label{ffh}.
\ee
\item
 For the case of pions, kaons and $\Lambda$, when SU(2)
invariance can be used this simplifies to
\be
R_{p-n}^{\pi^+ + \pi^-} = 2 D_u^{\pi^++\pi^-}(z,Q^2)
 =2 D_d^{\pi^++\pi^-}(z,Q^2).\label{ffpi}
\ee
\item
For $K$ mesons and $\Lambda$ hyperons,
\be
R_{p-n}^{K, \Lambda+\bar\Lambda} = 2 D_u^{K,
\Lambda+\bar\Lambda}(z,Q^2) = 2 D_d^{K,
\Lambda+\bar\Lambda}(z,Q^2) \label{ffK},
 \ee
  where the superscript $K$
 stands for the sum over all produced kaons:
\be
 K\equiv K^+ + K^- +K^0 +\bar K^0.
\ee
\end{itemize}

It will also be of great interest to compare these FF's with those
obtained from $e^+e^- \to hadrons$
 \cite{Binnewies,Kretzer} and those used in Monte Carlo models.

\vspace{1cm}

{\bf 2.} Given that $u_V(x,Q^2)$ and $d_V(x,Q^2)$ are well
determined from inclusive DIS one can proceed further to obtain
the other combinations of FF's $D_q^h - D_q^{\bar h} \equiv
D_q^{h-\bar h}$.
\begin{itemize}
\item
 One finds for $\pi^\pm$
 \be
D_u^{\pi^+ -\pi^- } &=&\frac{9\tilde\sigma_p^{\pi^+
-\pi^-}}{4u_V-d_V} =\frac{18\left( F_{1}^p\right)_{LO} R_p^{\pi^+
-\pi^-}}{4u_V-d_V}\nn\\ &=&\frac{9\tilde\sigma_n^{\pi^+
-\pi^-}}{4d_V-u_V} = \frac{18\left( F_{1}^n\right)_{LO} R_n^{\pi^+
-\pi^-}}{4d_V-u_V} \label{alpha}
 \ee
 Combined with (\ref{ffpi}) we
have expressions for $D_u^{\pi^+}$ and $D_u^{\pi^-}$ separately.
\item
For $K^\pm$ one obtains
\be
4D_u^{K^+ - K^- } - D_d^{K^+-K^-}
&=&\frac{9\left\{\tilde\sigma_p^{K^+ -K^-} - \tilde\sigma_n^{K^+
-K^-} \right\}}{u_V-d_V}\nn\\ &=&\frac{18\left[\left(
F_{1}^p\right)_{LO} R_p^{K^+ -K^-} -
 \left(F_{1}^n\right)_{LO} R_n^{K^+ -K^-}\right]}{u_V -d_V}\label{gamma}
\ee
 It is usually {\it assumed}, and this is presumably a very
good approximation,
 that
\be
D_d^{K^+} = D_d^{K^-}\label{delta}
\ee
in which case (\ref{gamma}) can be read as an expression for
$D_u^{K^+ - K^- }$. It is not possible to test relation
(\ref{delta}) without taking $s=\bar s$ and/or $\Delta s =
\Delta\bar s$, but such an
 approach is hard to justify given that any failure of
 (\ref{delta}) is presumably very small.
\item
For isoscalar hadrons like $\Lambda,\bar\Lambda$
\be
D_u^{\Lambda -\bar\Lambda } = D_d^{\Lambda -\bar\Lambda } =
\frac{6\left[\left( F_{1}^p\right)_{LO} R_p^{\Lambda^+ -\Lambda^-}
- \left( F_{1}^n\right)_{LO} R_n^{\Lambda^+
-\Lambda^-}\right]}{u_V -d_V}. \label{e} \ee
\end{itemize}
 Given that $\Delta u_V$ and $\Delta d_V$ are determined via
(\ref{A}) we can write analogous expression for the above $D_q^h$
functions using
 the polarized data.

Of course all the expressions (\ref{ffh}), (\ref{ffpi}),
(\ref{ffK}), (\ref{alpha}), (\ref{gamma}) and (\ref{e}), being LO
results, must be tested by demonstrating that the RH sides are
essentially independent of
 the passive variable $x$.

Now that we have determined $D_u^{\pi^++\pi^-}$ we can determine
 $D_s^{\pi^++\pi^-}$ in LO via
\be
 D_s^{\pi^++\pi^-}= \frac{9 \left(F_1^p+F_1^n\right)_{LO}
R_{p+n}^{\pi^++\pi^-} - 5q_+ D_u^{\pi^++\pi^-}} { 2\,(s+\bar s)}.
\ee

Similarly, since $D_u^{\Lambda +\bar\Lambda }$ is determined via
(\ref{ffK}), we can find $D_s^{\Lambda +\bar\Lambda }$ from
\be
 D_s^{\Lambda +\bar\Lambda }= \frac{9
\left(F_1^p+F_1^n\right)_{LO} R_{p+n}^{\Lambda +\bar\Lambda } -
5q_+ D_u^{\Lambda +\bar\Lambda }} { 2\,(s+\bar s)}.
 \ee


\subsection{The non-strange sea quark densities revisited,  in LO}


Now that we have determined $ D_u^{\pi^++\pi^-}$ and
 $D_s^{\pi^++\pi^-}$ in LO we can, in principle, determine
$\Delta q_+(x,Q^2)$ and $\Delta s(x,Q^2)+ \Delta\bar s(x,Q^2)$
from the semi-inclusive and inclusive relations, in LO,
\be
g_1^p +g_1^n& =& \frac{5\Delta q_+ + 2( \Delta s + \Delta \bar
s)}{18} \\
\Delta A_{p+n}^{\pi^++\pi^-}& =& \frac{5\Delta q_+ D_u^{\pi^++\pi^-}
 + 2 ( \Delta s + \Delta \bar s) D_s^{\pi^++\pi^-} } {5 q_+
D_u^{\pi^++\pi^-} + 2\,( s+\bar s) D_s^{\pi^++\pi^-} }.
\ee

Such a determination of $\Delta q_+$ in LO is likely to be
reliable, but $ \Delta s + \Delta\bar s$ and the individual
 $\Delta \bar u$ and $\Delta \bar d$ obtained in LO via
(\ref{sea2}) may be subject to significant uncertainty.

We note that there is an alternative way to determine $\Delta
q_+$, but it requires the ability to detect $K^0$. In that case
one has, in LO,
\be
\Delta q_+ = q_+
\frac{\Delta\tilde\sigma_p^{K^+ + K^- -(K^0 +\bar K^0)} +
\Delta\tilde\sigma_n^{K^+ + K^- -(K^0 +\bar K^0)}}
{\tilde\sigma_p^{K^+ + K^- -(K^0 +\bar K^0)} +
\tilde\sigma_n^{K^+ + K^- -(K^0 +\bar K^0)}}.
\ee


\subsection{The strange quark density $ \Delta s + \Delta\bar s$ in LO}


The approach to $ \Delta s + \Delta\bar s$ in Section 5.3 and the
approach discussed
 in \cite{we} are unlikely to be reliable. The problem is that
for production of pions the strange quark contribution is "doubly
small", since e.g. one must compare $\Delta uD_u^\pi$ with
$\Delta sD_s^\pi$ in which both $\vert \Delta s + \Delta\bar
s\vert \ll \vert\Delta u\vert$ and $\vert D_s^\pi\vert\ll\vert
D_u^\pi\vert$. For kaons and $\Lambda$ hyperons it is somewhat
better in that $\vert D_u^K\vert\approx\vert D_s^K\vert$ and
 $\vert D_u^{\Lambda +\bar\Lambda}\vert\ =
\vert D_d^{\Lambda +\bar\Lambda}\vert\approx
\vert D_s^{\Lambda +\bar\Lambda}\vert$, but this is similar
to the situation in inclusive DIS where we know that the LO
determination of $ \Delta s + \Delta\bar s$ is quite unreliable.

The only possibility we can see for a reasonable determination of
$\Delta s$ in LO is via $\phi$ production. For in this case one
has, $\vert \Delta s + \Delta\bar s\vert \ll \vert\Delta u\vert$
 but presumably $\vert D_s^\phi\vert\gg\vert D_u^\phi\vert$ so
 that the strange and non-strange quarks are on equal footing.

One has, by charge conjugation invariance $D_s^\phi = D_{\bar
s}^\phi$, and it should be quite safe to take $D_u^\phi = D_{\bar
u}^\phi
 =D_d^\phi = D_{\bar d}^\phi$. One then obtains in LO \be \frac{
 \Delta s + \Delta\bar s}{s+\bar s} =
\frac{3\Delta\tilde\sigma_{p+n}^\phi -
 5\left(\frac{\Delta q_+}{\Delta
q_3}\right)\Delta\tilde\sigma_{p-n}^\phi}{3\tilde\sigma_{p+n}^\phi
- 5\left(\frac{ q_+}{
q_3}\right)\tilde\sigma_{p-n}^\phi}.\label{deltas1} \ee Moreover
one has expressions for the fragmentation functions as well:
\be
D_s^\phi &=& \frac{3}{2(s+\bar s)}
\left\{3\tilde\sigma_{p+n}^\phi -
5\left(\frac{q_+}{q_3}\right)\tilde\sigma_{p-n}^\phi\right\}\label{deltas2}\\
D_u^\phi &=& \frac{3\tilde\sigma_{p-n}^\phi}{q_3}.\label{deltas3}
\ee
As always expressions (\ref{deltas1}) - (\ref{deltas3}) must be
tested
 for non-dependence on the relevant passive variable.

Finally we note that the same eqs. (\ref{deltas1}) -
(\ref{deltas3})
 hold for $K$, $\Lambda+\bar\Lambda$ and $\pi$-production if the
superscript $\phi$ is replaced by $K$, $\Lambda +\bar\Lambda$ and
$\pi^+ + \pi^-$, respectively.
  And though the production rates are higher, the sensitivity to
the strange quarks in the $K$, $\Lambda+\bar\Lambda$ and
$\pi$-productions is lower.

\subsection{Concerning $s=\bar s$ and $ \Delta s =  \Delta \bar s$ in
 LO
\protect\footnote{We are grateful to M. Anselmino, M. Boglione and
U.D'Alesio for
 drawing our attention to this issue}}

In the analysis of DIS it is conventional to assume $s=\bar s$
and $\Delta s = \Delta\bar s$. However there are models and
arguments \cite{a}
 which suggest that these equalities might not hold.

Within the limitations of the LO we can test these relationships
via
 $(K^+,K^-)$ and $(\Lambda ,\bar\Lambda )$ production. One has
for the unpolarized case, assuming $D_d^{K^+ -K^-}=0$ several
different possibilities:
\be
(s-\bar s) D_s^{K^+-K^-}& =&
18\,\left(F_1^p\right)_{LO} R_p^{K^+ -K^-} -
 4u_V D_u^{K^+ -K^-}\nn\\ &=&18\, \left(F_1^n\right)_{LO}
R_n^{K^+ -K^-} -
 4d_V D_u^{K^+ -K^-}\nn\\ &=&\frac{9\left\{u_V\tilde\sigma_n^{K^+
-K^-} - d_V \tilde\sigma_p^{K^+ -K^-}\right\}}
{u_V-d_V}.\label{e1}
 \ee
Then for the polarized case, given that $\Delta u_V$ and
 $\Delta d_V$ are known from (\ref{A}) and $(s-\bar s)D_s^{K^+
-K^-}$ is determined, we can proceed to determine $(\Delta s -
\Delta\bar s) D_s^{K^+ -K^-} $ from $\Delta A_p^{K^+ -K^-}$ or
$\Delta A_n^{K^+ -K^-}$:
\be
\Delta A_p^{K^+ -K^-} =
\frac{4\Delta u_V D_u^{K^+ -K^-} +
(\Delta s-\Delta\bar s)D_s^{K^+ -K^-}} {4 u_V D_u^{K^+ -K^-} +
(s-\bar s)D_s^{K^+ -K^-}}\nn\\
\Delta A_n^{K^+ -K^-}=
\frac{4\Delta d_V D_u^{K^+ -K^-} +
(\Delta s-\Delta\bar s)D_s^{K^+ -K^-}} {4 d_V D_u^{K^+ -K^-} +
(s-\bar s)D_s^{K^+ -K^-}},
\label{e2}
\ee
$D_u^{K^+-K^-}$ is assumed to be known through (\ref{gamma}).

For isoscalar hadrons like $\Lambda ,\bar\Lambda$ we have the
possiblities
\be
(s-\bar s) D_s^{\Lambda -\bar\Lambda }
&=&18\,\left(F_1^p\right)_{LO} R_p^{\Lambda -\bar\Lambda} -
 (4u_V + d_V) D_u^{\Lambda -\bar\Lambda }\nn\\
&=&18\,\left(F_1^n\right)_{LO} R_n^{\Lambda -\bar\Lambda} -
 (4d_V + u_V) D_u^{\Lambda -\bar\Lambda }\nn\\ &=&
\frac{3}{u_V-d_V}\left\{ \left( 4u_V+d_V\right)
 \tilde\sigma_n^{\Lambda -\bar\Lambda} -\left(4d_V+u_V\right)
 \tilde\sigma_p^{\Lambda -\bar\Lambda}\right\},\label{e3}
\ee
 where $D_u^{\Lambda -\bar\Lambda }$ is assumed to be
determined in (\ref{e}). Here and in eq. (\ref{e1}) it is clear
that $\tilde\sigma^{K^+-K^-}_N$ and $\tilde\sigma^{\Lambda -
\bar\Lambda}_N$ can be replaced by the corresponding
experimentally measured quantities $2\left( F_{1}^N\right)_{LO}
R_N^{K^+-K^-} $ or $2\left( F_{1}^N\right)_{LO} R_N^{\Lambda
-\bar\Lambda} $. Then $(\Delta s - \Delta\bar s) D_s^{\Lambda
-\bar\Lambda } $ can be determined via $\Delta A_p^{\Lambda
-\bar\Lambda }$ or $\Delta A_n^{\Lambda -\bar\Lambda }$:
\be
\Delta A_p^{\Lambda -\bar\Lambda } =
\frac{(4\Delta u_V + \Delta d_V) D_u^{\Lambda -\bar\Lambda } +
(\Delta s-\Delta\bar s)D_s^{\Lambda -\bar\Lambda }} {(4 u_V +
d_V) D_u^{\Lambda -\bar\Lambda } + (s-\bar s)D_s^{\Lambda
-\bar\Lambda }}\nn\\
\Delta A_n^{\Lambda -\bar\Lambda } =
\frac{(4\Delta d_V + \Delta u_V) D_u^{\Lambda -\bar\Lambda } +
(\Delta s-\Delta\bar s)D_s^{\Lambda -\bar\Lambda }} {(4 d_V +
u_V) D_u^{\Lambda -\bar\Lambda } + (s-\bar s)D_s^{\Lambda
-\bar\Lambda }}
\label{e4}
\ee

Although (\ref{e1}), (\ref{e2}), (\ref{e3}) and (\ref{e4}), being
LO expressions, cannot be expected to yield accurate values for
$s-\bar s$ and $\Delta s-\Delta\bar s$,
 they should nonetheless enable one to say whether they are
compatible with zero since $D_s^{\Lambda -\bar\Lambda}$ should be
relatively large. Note that $s-\bar s$ and/or $\Delta
s-\Delta\bar s$ different from zero would break the independence
of the RH sides of (\ref{e2}) and (\ref{e4}) on the passive
variable $z$.
 Of particular interest is the speculation that $\Delta s \approx
-\Delta\bar s$ but $s\approx \bar s$, the consistency of which
could perhaps be
 tested from (\ref{e1}), (\ref{e2}), (\ref{e3}) and (\ref{e4}).


\section{Semi-inclusive analysis in NLO QCD}


The situation in NLO \cite{deFlorian,Graudenz} is much more
complicated than in LO, since factorization is replaced by
convolution, and it is also more complicated than inclusive DIS in
NLO since here one has to contend with double convolutions of the
form $q\otimes C\otimes D$ and $\Delta q\otimes \Delta C\otimes D$
for the unpolarized and polarized cases respectively, where $C$
and $\Delta C$ are Wilson coefficients first derived
 in \cite {Graudenz} and \cite{deFlorian}.

The double convolution is defined as
\be
q\otimes C\otimes D = \int{dx'\over x'}\int{dz'\over z'}
 q\left({x\over x'}\right) C( x',z') D\left({z\over z'}\right)
\ee
where the range of integration is given as follows:
\begin{itemize}
\item
${\bf I_1:}$ The range is
\be
{x\over x+(1-x)z} \leq x'\leq 1\quad with \quad z\leq z'\leq
1\label{conv}
\ee
\item
${\bf I_2:}$ In addition to
 (\ref{conv}) there is  the range
\be
x\leq x'\leq {x\over x+(1-x)z}\quad with\quad
\frac{x(1-x')}{x'(1-x)}\leq z'\leq 1.
\ee
\end{itemize}
Note, that contrary to the case of the usual DIS convolution, the
double convolution $q\otimes C\otimes D$ is not commutative.

We shall frequently encounter expressions of the form
\be
qD + \frac{\alpha_s}{2\pi} q\otimes C\otimes D \ee corresponding
to the LO plus NLO corrections. The flavour structure of the
 results becomes much more transparent if we adopt the following
symbolic notation:
\be
qD + \frac{\alpha_s}{2\pi} q\otimes C\otimes D = q\left[
1+\otimes \frac{\alpha_s}{2\pi} C\otimes \right] D.
\ee
Then the semi-inclusive polarized cross section
$\Delta\tilde\sigma_p^h$ defined in (\ref{Deltasigma}) is given
by
\be
\Delta\tilde\sigma_p^h& =&\sum_i e_i^2 \Delta q_i\left[ 1+
\otimes \frac{\alpha_s}{2\pi}
\Delta C_{qq}\otimes \right] D_{q_i}^h +\nn\\
&&+\left(\sum_i e_i^2 \Delta q_i\right) \otimes
\frac{\alpha_s}{2\pi}\Delta C_{qg}
\otimes D_G^h +\Delta  G\otimes \frac{\alpha_s}{2\pi} \Delta C_{gq}\otimes
 \left(\sum_i e_i^2 D_{q_i}^h\right)\,\label{deltasigmaNLO}
\ee
where the sum is over quarks and antiquarks of flavour $i$ and
parton
 densities and fragmentation functions are to be taken in NLO.

For the unpolarized semi-inclusive cross section in NLO it is not
possible to completely factor out the $y$-dependence.
Consequently $\tilde\sigma^h$ defined in (\ref{sigmah}) will
depend upon $y$ in NLO, in contrast to the LO situation in
(\ref{sigmaLO}).

In the notation of the seminal paper of Graudenz \cite{Graudenz}
the cross section is a sum of what he refers to as ``metric''
($M$) and ``longitudinal''
 ($L$) terms, with corresponding Wilson coefficients $C^M$ and
$C^L$. A further complication is that the Wilson coefficients are
different in the
 two regions of integration $I_1$ and $I_2$. Thus we have
coefficients $C^{jM}$, $C^{jL}$ with $j=1,2$.

We then define the $y$-dependent combinations of Wilson
coefficients:
\be
&& {\mathbb C}_{qq}^j = C_{qq}^{jM} +\left[1+4\gamma (y)\right]
 C_{qq}^{jL}\\ && {\mathbb C}_{qg}^j = C_{qg}^{jM}
+\left[1+4\gamma (y)\right]
 C_{qg}^{jL}\\ && {\mathbb C}_{gq}^j = C_{gq}^{jM}
+\left[1+4\gamma (y)\right]
 C_{gq}^{jL}
\ee
where
\be
\gamma (y) = \frac{1-y}{1+(1-y)^2}.
\ee

Then the unpolarized semi-inclusive cross section can be written
in a form analogous to the polarized one:
\be
\tilde\sigma_p^h & =&
\sum_i e_i^2 q_i\left[ 1+\otimes \frac{\alpha_s}{2\pi}
 {\mathbb C}_{qq}\otimes \right] D_{q_i}^h +\nn\\ &&+\left(\sum_i
e_i^2 q_i\right) \otimes \frac{\alpha_s}{2\pi} {\mathbb C}_{qg}
\otimes D_G^h + G\otimes \frac{\alpha_s}{2\pi} {\mathbb C}_{gq}\otimes
 \left(\sum_i e_i^2 D_{q_i}^h\right).\label{sigmaNLO}
\ee

In (\ref{sigmaNLO}) we have used the symbolic notation:
\be
q_i\otimes \frac{\alpha_s}{2\pi}
 {\mathbb C}_{qq}\otimes  D_{q_i}^h =
\int_{I_1} q_i\otimes \frac{\alpha_s}{2\pi}
 {\mathbb C}_{qq}^1\otimes D_{q_i}^h + \int_{I_2} q_i\otimes
\frac{\alpha_s}{2\pi}
 {\mathbb C}_{qq}^2\otimes  D_{q_i}^h\,,
\ee
and analogously for ${\mathbb C}_{qg}$ and ${\mathbb C}_{gq}$.

Note that in NLO the unpolarized inclusive cross section,
$\tilde\sigma^{DIS}$ (\ref{sigmaDIS}) is given by
\be
\tilde\sigma^{DIS}= 2F_1\left[ 1 + 2\gamma (y) R\right],
\ee
where $R$ is the usual DIS ratio of longitudinal to transverse
cross sections. Note that strictly speaking here and throughout
the rest of this
 paper $R$ should be replaced by $R\longrightarrow \left(
R-\frac{4M^2x^2}{Q^2}\right)/\left(1+\frac{4M^2x^2}{Q^2}\right)$,
 since the correction terms may be important for low values of
$Q^2$.

As in the LO discussion we doubt the reliability of a global NLO
analysis of inclusive and semi-inclusive data, and we suggest
that one should feed into
 the semi-inclusive analysis as much reliable information as one
 can from other sources.

As we shall see there is the oft found opposition between what is
 simple theoretically and what is simple experimentally. However,
if the systematic errors in detection efficiencies
 can be brought under control then we can make remarkable
 theoretical simplifications and we can then extract a vast
 amount of information from the semi-inclusive data. This is an
 important experimental challenge as will be seen from the power
 of the results given below.


\subsection{Simplification of the semi-inclusive NLO results}


Bearing in mind the NLO result (\ref{g1p}) for $g_1^p$ for
polarized DIS and
 the analogous result for $F_1$ in unpolarized DIS, we see that
 in the second term of (\ref{sigmaNLO}) we may make the
 replacement, correct to NLO,
\be
\sum_i e_i^2 q_i \longrightarrow 2 F_1^p
\ee
 and similarly, in the analogous eq. for $\Delta\tilde\sigma_p$
in (\ref{deltasigmaNLO}):
\be
\sum_i e_i^2 \Delta q_i \longrightarrow 2 g_1^p.
\ee
Note that here {\it and throughout the rest of this paper}
$F_1^{p,n}$ and $g_1^{p,n}$ are the experimentally measured
 structure functions.

Further in the reaction $e^+(p_+) + e^-(p_-) \to h +X$ in the
kinematic region
 when we can neglect $Z^0$-exchange effects, we have
\be
\sigma^h(z,\cos\theta ,Q^2) = \frac{3}{8} (1+\cos^2\theta )
 \sigma_T^h(z,Q^2) + \frac{3}{4} (1-\cos^2\theta )
 \sigma_L^h(z,Q^2)
\ee
where $Q^2 =(p_+ + p_-)^2$.

In NLO QCD one has
\be
\sigma_T^h(z,Q^2) & =&3\sigma_0\left\{\sum_i e_i^2 D_{q_i}^h\otimes
\left( 1+ \frac{\alpha_s}{2\pi}
 C_{q}^T\right)\right.+\nn\\ &&\hspace{1cm}+\left.\sum_i e_i^2
D_G^h \otimes \frac{\alpha_s}{2\pi} C_{G}^T\right\}
\ee
where the sum is over quarks and antiquarks, the $C^T$'s are
Wilson
 coefficients, and
\be
\sigma_0 =\frac{4\pi\alpha^2}{3Q^2}.
\ee

Then, correct to the required NLO accuracy, in the third term in
 (\ref{sigmaNLO}), and in its polarized analogue
 (\ref{deltasigmaNLO}), we may make the replacement
\be
\sum_i e_i^2 D_{q_i}^h = \frac{\sigma_T^h(z,Q^2)}{3\sigma_0},
\ee
where $\sigma^h_T$ is the experimentally measured cross section.

Hence, for the unpolarized and polarized semi-inclusive cross
sections,
 in NLO accuracy, we have
\be
\tilde\sigma_p^h& =
&\sum_i e_i^2 q_i\left[ 1+\otimes \frac{\alpha_s}{2\pi}
 {\mathbb C}_{qq}\otimes \right] D_{q_i}^h +\nn\\ &&+ 2 F_1^p
\otimes \frac{\alpha_s}{2\pi} {\mathbb C}_{qg}
\otimes D_G^h +
 \frac{1}{3\sigma_0} G\otimes \frac{\alpha_s}{2\pi} {\mathbb
C}_{gq}\otimes
 \sigma_T^h\label{sigmaNLOnew}
\ee
\be
\Delta \tilde\sigma_p^h& =&\sum_i e_i^2 \Delta q_i\left[ 1+
\otimes \frac{\alpha_s}{2\pi}
\Delta C_{qq}\otimes \right] D_{q_i}^h +\nn\\
&&+ 2 g_1^p \otimes \frac{\alpha_s}{2\pi}\Delta C_{qg}
\otimes D_G^h +
 \frac{1}{3\sigma_0} \Delta G\otimes \frac{\alpha_s}{2\pi}\Delta
 C_{gq}
\otimes\sigma_T^h.\label{deltasigmaNLOnew}
\ee
Given that the unpolarized gluon density is reasonably well
known, the last term in (\ref{sigmaNLOnew}) can be considered as
a known quantity. In the following we take as known quantities
the NLO values for $q_+(x,Q^2)$, $q_3(x,Q^2)$, $u_V(x,Q^2)$,
$d_V(x,Q^2)$, $G(x,Q^2)$ and $\Delta q_3(x,Q^2)$.


\subsection{The polarized valence densities in NLO}


Using charge conjugation invariance one obtains, for
semi-inclusive pion
 production
\be
&&R_p^{\pi^+-\pi^-} = \frac{[4u_V -d_V][1+\otimes (\alpha_s/2\pi)
{\mathbb C}_{qq}\otimes ] D_u^{\pi^+ -\pi^-}}{18F_1^p\,[1+2\gamma
(y)\,R^{\,p}]}\nn\\ &&R_n^{\pi^+-\pi^-} =\frac{[4d_V
-u_V][1+\otimes (\alpha_s/2\pi) {\mathbb C}_{qq}\otimes ]
D_u^{\pi^+ -\pi^-}}{18F_1^p\,[1+2\gamma
(y)\,R^{\,p}]}.\label{DuNLO} \ee The only unknown function in
these expressions is $D_u^{\pi^+ -\pi^-}(z,Q^2)$, which evolves as
a non-singlet and does not mix with other FF's. A $\chi^2$
analysis of either or both of the equations (\ref{DuNLO}) should
thus determine
 $D_u^{\pi^+ -\pi^-}$ in NLO without serious ambiguity.

Assuming $D_u^{\pi^+ -\pi^-}$ is now known, one can then
determine $\Delta u_V$ and $\Delta d_V$ in NLO via the equations
\be
\Delta A_p^{\pi^+ -\pi^-}
= \frac{ (4\Delta u_V -\Delta d_V)
[1+ \otimes (\alpha_s/2\pi) \Delta C_{qq}\otimes ] D_u^{\pi^+
-\pi^-}} {(4u_V -d_V)[1+\otimes (\alpha_s /2\pi) {\mathbb
C}_{qq}\otimes ] D_u^{\pi^+ -\pi^-}}\label{valence1NLO}
\ee
\be
\Delta A_n^{\pi^+ -\pi^-}
= \frac{ (4\Delta d_V -\Delta u_V)
[1+ \otimes (\alpha_s /2\pi) \Delta C_{qq}\otimes ] D_u^{\pi^+
-\pi^-}} {(4d_V -u_V)[1+\otimes (\alpha_s /2\pi) {\mathbb
C}_{qq}\otimes ] D_u^{\pi^+ -\pi^-}}\label{valence2NLO}
\ee
where, of course, $\Delta u_V$ and $\Delta d_V$ evolve as
non-singlets and do
 not mix with other densities. Eqs. (\ref{valence1NLO}) and
 (\ref{valence2NLO}) determine the densities $\Delta u_V$ and
 $\Delta d_V$ in NLO without any assumptions about the less known
 polarized gluon and sea densities.


\subsection{SU(2) symmetry of the  sea quark densities in NLO}


Once $\Delta u_V$ and $\Delta d_V$ are known in NLO we can
calculate
\be
[\Delta\bar u(x,Q^2) - \Delta\bar d(x,Q^2)]_{NLO} =
\frac{1}{2}[\Delta q_3(x,Q^2) + \Delta d_V(x,Q^2) - \Delta u_V(x,Q^2)]_{NLO},
\label{seaNLO}
\ee
Eq. (\ref{seaNLO}) determines the breaking of SU(2) symmetry for
the
 polarized sea densities in NLO without requiring any knowledge
of $\Delta \bar q$ and $\Delta G$. It will be interesting to
 compare the values obtained from (\ref{seaNLO}) with information
on $\Delta \bar u(x)$ and $\Delta\bar d(x)$ which will emerge
from Drell-Yan and $W^\pm$ production experiments at
RHIC~\cite{Bourelly}.

The separate determination of $\Delta \bar u$ and $\Delta\bar d$
requires knowledge of $\Delta q_+$, defined in (\ref{deltaq+}),
 in NLO.

Note that determining, say, $\Delta \bar u$ via $\Delta\bar u =
\frac{1}{2} [(\Delta u + \Delta\bar u) -\Delta u_V]$ is
unreliable, since the
 determination of $(\Delta u + \Delta\bar u)$ from {\it
 inclusive} DIS involves knowledge of gluon and strange quark
 densities.

As in the LO case we can only determine $\Delta q_+$ after
obtaining some
 information about the fragmentation functions.


\subsection{Fragmentation functions in NLO}


We consider the sum for the unpolarized production of $\pi^+$ and
$\pi^-$. We have
\be
R_{p-n}^{\pi^++\pi^-} =
\frac{q_3\left\{\left[ 1+\otimes \frac{\alpha_s}{2\pi}
 {\mathbb C}_{qq}\otimes \right] D_{u}^{\pi^++\pi^-} + \otimes
 \frac{\alpha_s}{2\pi} {\mathbb C}_{qg}
\otimes D_G^{\pi^++\pi^-}\right\}}
{6\left[F_1^p\left(1+2\gamma (y) R^p\right) - F_1^n \left(
1+2\gamma (y) R^n\right)\right] }
\label{q3NLO}
\ee
\be
\Delta A_{p-n}^{\pi^++\pi^-} =
\frac{\Delta q_3\left\{\left[ 1+\otimes \frac{\alpha_s}{2\pi}
 \Delta C_{qq}\otimes \right] D_{u}^{\pi^++\pi^-} + \otimes
 \frac{\alpha_s}{2\pi} \Delta C_{qg}
\otimes D_G^{\pi^++\pi^-}\right\}}
{ q_3\left\{\left[ 1+\otimes \frac{\alpha_s}{2\pi}
  {\mathbb C}_{qq}\otimes \right] D_{u}^{\pi^++\pi^-} + \otimes
 \frac{\alpha_s}{2\pi} {\mathbb C}_{qg}
\otimes D_G^{\pi^++\pi^-}\right\}}\, \cdot\label{deltaq3NLO}
\ee
The only unknown functions in these relations are
$D_u^{\pi^++\pi^-}$ and $D_G^{\pi^++\pi^-}$, which will mix with
each other under evolution. Thus it should be possible to obtain
them from a $\chi^2$ analysis of
 (\ref{q3NLO}) and (\ref{deltaq3NLO}).

Once we know $D_u^{\pi^++\pi^-}$ and utilise $D_u^{\pi^+-\pi^-}$
from Section 6.2 we clearly have access to both $D_u^{\pi^+}$ and
$D_u^{\pi^-}$.

Note that if $K^0$ can be detected one can obtain information on
$D_u^{K}$ and $D_G^{K}$, where $K= K^++K^-+K^0+\bar K^0$.
 One simply replaces the labels $\pi^+ + \pi^-$ by $K$ everywhere
in (\ref{q3NLO}) and (\ref{deltaq3NLO}). Analogous equations hold
also if $\pi^+ + \pi^-$ is replaced by $\Lambda +\bar\Lambda $,
if $\Lambda$
 is detected.

Returning to the case of $\pi^+ + \pi^-$, the ratio
$R_{p+n}^{\pi^++\pi^-}$ allows the determination of the only
unknown function $D_s^{ \pi^+ + \pi^-}$. We have
\be
&R_{p+n}^{\pi^++\pi^-} =\nn&\\ &= \left\{\left(5q_+\left[1+
\otimes  \frac{\alpha_s}{2\pi} {\mathbb C}_{qq}\otimes \right]
 D_u^{ \pi^+ + \pi^-} + 2(s+\bar s)\left[1+
\otimes  \frac{\alpha_s}{2\pi} {\mathbb C}_{qq}\otimes \right]
D_s^{ \pi^+ + \pi^-}\right)\right. +
\nn&\\
&\hspace{0.5cm}+ 18 (F_1^p + F_1^n) \otimes
 \frac{\alpha_s}{2\pi}{\mathbb C}_{gq}\otimes D_G^{ \pi^+ +
 \pi^-} + \left.\frac{6}{\sigma _0}G\otimes \frac{\alpha_s}{2\pi}
 {\mathbb C}_{gq}\otimes
\sigma_T^{ \pi^+ + \pi^-}\right\}/\nn&\\
&\hspace{0.5cm}/ 6\left[F_1^p\left(1+2\gamma (y) R^p\right) +
F_1^n \left( 1+2\gamma (y) R^n\right)\right] &
\ee
 Under evolution $D_u^{ \pi^+ + \pi^-}$ and $D_s^{ \pi^+ +
 \pi^-}$ mix with $D_G^{ \pi^+ + \pi^-}$, but this is not a
 problem since the latter is supposed to be known.
It would be interesting to compare these results with those from
$e^+e^-\to hadrons$, obtained recently in \cite{Kretzer}.

 Again analogous sets of equations holds
for kaon and $\Lambda ,\bar\Lambda $
 production. One simply replaces $\pi^++\pi^-$ by $K$ or $\Lambda
+\bar\Lambda$
 and this allows the determination of the only unknown function
$D_s^K$ or $D_s^{\Lambda +\bar\Lambda}$, respectively.


\subsection{The  sea-quark densities in NLO}


With the NLO knowledge of $D_u^{ \pi^+ + \pi^-}$, $D_s^{ \pi^+ +
\pi^-}$ and $D_G^{ \pi^+ + \pi^-}$ we are now in a position to try
to determine $\Delta q_+(x,Q^2)$, $\Delta s(x,Q^2) + \Delta\bar
s(x,Q^2)$ and $\Delta G(x,Q^2)$. We have, in NLO,
\be
g_1^p + g_1^n =\frac{1}{18} \left[5\Delta q_+
 + 2( \Delta s + \Delta\bar s)\right]
\otimes \left(1+
\frac{\alpha_s(Q^2)}{2\pi}\delta C_q\right)
+ \frac{\alpha_s(Q^2)}{2\pi}\Delta G \otimes \delta
C_G,\label{g1p+nNLO}
\ee
and
\be
\Delta  A_{p+n}^{ \pi^+ + \pi^-}=
\frac{\Delta \tilde\sigma^{\pi^++\pi^-}_p
+ \Delta \tilde\sigma^{\pi^++\pi^-}_n } {
\tilde\sigma^{\pi^++\pi^-}_p + \tilde\sigma^{\pi^++\pi^-}_n }
\ee
where
\be
&\Delta \tilde\sigma^{\pi^++\pi^-}_p + \Delta
\tilde\sigma^{\pi^++\pi^-}_n =\nn&\\ &=\left(5\Delta q_+\left[
1+\otimes \frac{\alpha_s}{2\pi}
\Delta  C_{qq}\otimes \right]  D_{u}^{ \pi^+ + \pi^-}
+ 2( \Delta s + \Delta\bar s)\left[ 1+\otimes
\frac{\alpha_s}{2\pi}
\Delta  C_{qq}\otimes \right] D_{s}^{ \pi^+ + \pi^-}\right)
 +\nn&\\ & + 18 \left( g_1^p + g_1^n\right)\otimes
 \frac{\alpha_s(Q^2)}{2\pi}\Delta C_{qg} \otimes D_G^{ \pi^+ +
\pi^-} + \frac{6}{\sigma_0} \Delta G
\otimes  \frac{\alpha_s(Q^2)}{2\pi}\Delta C_{gq}
 \otimes \sigma_T^{ \pi^+ + \pi^-}&
\ee
and
\be
&\tilde\sigma^{\pi^++\pi^-}_p + \tilde\sigma^{\pi^++\pi^-}_n
=\nn&\\ &= \left( 5 q_+ \left[ 1+\otimes \frac{\alpha_s}{2\pi}
{\mathbb C}_{qq}\otimes \right] D_{u}^{ \pi^+ + \pi^-} + 2(s+\bar
s)\left[ 1+\otimes \frac{\alpha_s}{2\pi} {\mathbb C}_{qq}\otimes
\right] D_{s}^{ \pi^+ + \pi^-} \right)
 +\nn&\\ & + 18 \left( F_1^p + F_1^n\right)\otimes
 \frac{\alpha_s(Q^2)}{2\pi} {\mathbb C}_{qg} \otimes D_G^{ \pi^+
+ \pi^-} + \frac{6}{\sigma_0} G
\otimes  \frac{\alpha_s(Q^2)}{2\pi} {\mathbb C}_{gq}
 \otimes \sigma_T^{ \pi^+ + \pi^-}\label{sigmap+nNLO}.&
\ee
Note that an analogous set of equations hold for $\pi^++\pi^-$
replaced by $K$ or $\Lambda +\bar\Lambda$.

Eqs. (\ref{g1p+nNLO}) to (\ref{sigmap+nNLO}) contain three
unknown functions $\Delta q_+$, $\Delta s+\Delta\bar s$ and
$\Delta G$, which nonetheless can all
 be determined in principle because of their different evolution
 in $Q^2$. However, to be at all efficacious such a determination
 would require a huge range of $Q^2$, far larger than is
 available in present day polarised DIS.

On the other hand there is a direct and superior method for
obtaining
 $\Delta G$, namely via $c\bar c$ production. This is one of the
major goals of the COMPASS experiment at CERN. We shall thus
assume that $\Delta G$ has been determined, so that the last term
on the RHS of (\ref{deltasigmaNLOnew}) may be taken to be known.

It should be then straightforward to determine $\Delta q_+$ and
 $\Delta s +\Delta\bar s$ in NLO from a $\chi^2$ analysis of
(\ref{g1p+nNLO}) to (\ref{sigmap+nNLO}) in which the evolution of
$\Delta q_+$ and $\Delta s +\Delta\bar s$ would involve mixing
with the supposed known $\Delta G$.

An independent determination of $\Delta q_+$ and
 $\Delta s + \Delta\bar s$ could be obtained by combining
(\ref{g1p+nNLO}) with $\Delta A_{p+n}^{K}$ for kaons and
 $\Delta A_{p+n}^{\Lambda +\bar\Lambda}$ for $\Lambda
,\bar\Lambda$ production.

Once $\Delta q_+$ is known in NLO, we can obtain the individual
 $\Delta \bar u$ and $\Delta\bar d$ from (\ref{seaNLO}).

\subsection{Concerning $s=\bar s$ and $\Delta s=\Delta\bar s$ in NLO}

It is possible to get some information on
 $s - \bar s$ and $ \Delta s - \Delta\bar s $ in NLO.

We have, assuming $D_d^{K^+} = D_d^{K^-}$,
\be
&R_p^{K^+-K^-} =&\nn\\ &= \left\{4u_V\left[ 1+\otimes
\frac{\alpha_s}{2\pi}
 {\mathbb C}_{qq}\otimes \right] D_{u}^{K^+-K^-} + (s - \bar
 s)\left[ 1+\otimes \frac{\alpha_s}{2\pi} {\mathbb C}_{qq}\otimes
 \right] D_{s}^{K^+-K^-}\right\}
/& \nn\\
&\hspace{1cm} / 18 F_1^p\left[1 + 2\gamma (y) R^p\right] &
\label{n2}
\ee
\be
&R_n^{K^+-K^-} =\nn&\\ &=\left\{4d_V \left[ 1+\otimes
\frac{\alpha_s}{2\pi}
 {\mathbb C}_{qq}\otimes \right] D_{u}^{K^+-K^-} + (s - \bar s)
 \left[ 1+\otimes \frac{\alpha_s}{2\pi} {\mathbb C}_{qq}\otimes
 \right] D_{s}^{K^+-K^-}\right\}
/\nn&\\
&\hspace{1cm}/ 18 F_1^n\left[1 + 2\gamma (y) R^n\right] &
\label{n3}
\ee
 These two equations, taken together with those for $\Delta
A_p^{K^+-K^-}$ and $\Delta A_n^{K^+-K^-}$:
\be
&\Delta A_p^{K^+-K^-}=\nn&\\ &=\left(4 \Delta u_V
\left[ 1+\otimes \frac{\alpha_s}{2\pi}
 \Delta C_{qq}\otimes \right] D_{u}^{K^+ -K^-} + ( \Delta s -
\Delta\bar s)\left[ 1+\otimes \frac{\alpha_s}{2\pi}
 \Delta C_{qq}\otimes \right] D_{s}^{K^+ -K^-}\right)/\nn&\\
&/\left(4 u_V \left[ 1+\otimes \frac{\alpha_s}{2\pi}
  {\mathbb C}_{qq}\otimes \right] D_{u}^{K^+ -K^-} + ( s - \bar
s)\left[ 1+\otimes \frac{\alpha_s}{2\pi}
  {\mathbb C}_{qq}\otimes \right] D_{s}^{K^+ -K^-}\right)&
\ee
\be
&\Delta A_n^{K^+-K^-}=\nn&\\ &= \left(4 \Delta d_V \left[
1+\otimes \frac{\alpha_s}{2\pi}
\Delta  C_{qq}\otimes \right]  D_{u}^{K^+ -K^-}+
 ( \Delta s - \Delta\bar s)\left[ 1+\otimes \frac{\alpha_s}{2\pi}
\Delta  C_{qq}\otimes \right] D_{s}^{K^+ -K^-}\right)/\nn&\\
&/\left(4 d_V \left[ 1+\otimes \frac{\alpha_s}{2\pi}
  {\mathbb C}_{qq}\otimes \right] D_{u}^{K^+ -K^-} + ( s - \bar
s)\left[ 1+\otimes \frac{\alpha_s}{2\pi}
  {\mathbb C}_{qq}\otimes \right] D_{s}^{K^+ -K^-}\right)&
\ee
provide four equations for the
 three unknown functions $(s-\bar s)\otimes D_s^{K^+-K^-}$, $(
\Delta s - \Delta\bar s)\otimes D_s^{K^+-K^-}$ and
 $D_u^{K^+-K^-}$, so that, in principle, all can be determined
 via a $\chi^2$ analysis.

In addition one has
\be
&R_p^{\Lambda -\bar\Lambda}=\nn&\\ &=\left\{(4u_V+d_V) \left[
1+\otimes \frac{\alpha_s}{2\pi}
 {\mathbb C}_{qq}\otimes \right] D_{u}^{\Lambda -\bar\Lambda} +
(s - \bar s)\left[ 1+\otimes \frac{\alpha_s}{2\pi}
 {\mathbb C}_{qq}\otimes \right] D_{s}^{\Lambda -\bar\Lambda}\right\}
/\nn&\\
&\hspace{1cm} / 18 F_1^p\left[1 + 2\gamma (y) R^p\right] &
\label{n4}
\ee
and
\be
&R_n^{\Lambda -\bar\Lambda}=\nn&\\ &=\left\{(4d_V+u_V)\left[
1+\otimes \frac{\alpha_s}{2\pi}
 {\mathbb C}_{qq}\otimes \right] D_{u}^{\Lambda -\bar\Lambda} +
(s - \bar s)\left[ 1+\otimes \frac{\alpha_s}{2\pi}
 {\mathbb C}_{qq}\otimes \right] D_{s}^{\Lambda -
\bar\Lambda}\right\}/\nn&\\
&\hspace{1cm} / 18 F_1^n\left[1 + 2\gamma (y) R^n\right] &
\ee
together with $\Delta A_p^{\Lambda -\bar\Lambda}$ and $\Delta
A_n^{\Lambda -\bar\Lambda}$:
\be
&\Delta A_p^{\Lambda -\bar\Lambda}=\nn&\\ &= \left((4 \Delta u_V
+\Delta d_V)
 \left[ 1+\otimes \frac{\alpha_s}{2\pi}
\Delta  C_{qq}\otimes \right]  D_{u}^{\Lambda -\bar\Lambda}
+ ( \Delta s - \Delta\bar s)\left[ 1+\otimes
\frac{\alpha_s}{2\pi}
\Delta  C_{qq}\otimes \right] D_{s}^{\Lambda -\bar\Lambda}\right)/\nn&\\
&/\left( (4 u_V +d_V) \left[ 1+\otimes \frac{\alpha_s}{2\pi}
 {\mathbb C}_{qq}\otimes \right] D_{u}^{\Lambda -\bar\Lambda} + (
s - \bar s)\left[ 1+\otimes \frac{\alpha_s}{2\pi}
 {\mathbb C}_{qq}\otimes \right] D_{s}^{\Lambda -\bar\Lambda}\right)&
\ee
\be
&\Delta A_n^{\Lambda -\bar\Lambda}=\nn&\\ &=\left( (4 \Delta d_V
+\Delta d_V)\left[ 1+\otimes \frac{\alpha_s}{2\pi}
\Delta  C_{qq}\otimes \right]
 D_{u}^{\Lambda -\bar\Lambda} + ( \Delta s - \Delta\bar s)\left[
1+\otimes \frac{\alpha_s}{2\pi}
\Delta  C_{qq}\otimes \right] D_{s}^{\Lambda -\bar\Lambda}\right)/\nn&\\
&/\left( (4 d_V+d_V) \left[ 1+\otimes \frac{\alpha_s}{2\pi}
 {\mathbb C}_{qq}\otimes \right] D_{u}^{\Lambda -\bar\Lambda} + (
s - \bar s) \left[ 1+\otimes \frac{\alpha_s}{2\pi}
 {\mathbb C}_{qq}\otimes \right]  D_{s}^{\Lambda -\bar\Lambda}\right)&
.\label{n8}
\ee
 These provide four more equations but only two new unknown
functions $D_u^{\Lambda -\bar\Lambda}$ and $D_s^{\Lambda
-\bar\Lambda}$. The system of eight equations (\ref{n2}) -
(\ref{n8}) therefore over-constrains the unknown functions and
might, hopefully, allow a reasonable determination of the
relation between $s-\bar s$ and $\Delta s - \Delta \bar s$ and
whether or not $s(x)=\bar s(x)$ and/or $\Delta s (x) = \Delta
\bar s(x)$. The actual determination of $s-\bar s$ or $\Delta s
-\Delta\bar s$ requires knowledge of either $D_s^{\Lambda
-\bar\Lambda}$ or $D_s^{K^+-K^-}$. These could be taken from the
study of $e^+e^-\to hX$.

This completes the determination of all the polarized densities
in NLO.

\section{Conclusions}

We have argued that the present LO QCD method of analysing
polarized semi-inclusive DIS, using the concept of purity, is
quite unjustified. We have also argued against attempts at a
global analysis, either in LO or in NLO QCD, based on the
combined data on polarized inclusive and semi-inclusive DIS and
taking as known exactly the relevant
 FF's.

Instead, we have presented a strategy for a step by step
evaluation of the polarized parton densities and fragmentation
functions from semi-inclusive data using selectively chosen
information from inclusive
 DIS reactions.

In our approach the usually made simplifying assumptions about
 relations between $\Delta\bar u$ and $\Delta\bar d$, and between
the strange and non-strange polarized sea densities
 are unnecessary and we have even considered the possibility that
$s(x)\not=\bar s (x)$ and $\Delta s(x)\not=\Delta\bar s (x)$.

Given the simplicity of the LO QCD analysis, we discuss where and
when it is likely to be reliable, and stress the need to test the
consistency of
 the LO treatment at each step. In this connection we have
introduced the concept of a {\it passive} variable in the
experimentally measured observables.
 We have also suggested how one might estimate the errors induced
in doing the LO analysis.

In the NLO treatment we have shown how the expressions for the
experimental
 observables can be much simplified by incorporating information from
the reaction $e^+ e^- \to hX$ in a novel way. Determination of the
polarized valence quark densities $\Delta u_V$ and $\Delta d_V$ is
shown to be relatively straight forward, as is the difference
$\Delta\bar u - \Delta\bar d$. However we argue that the
determination of $\Delta\bar u$, $\Delta\bar d$, $\Delta s
+\Delta\bar s$
 and $\Delta G$ separately, from semi-inclusive DIS involving
production of $\pi$, $K$, $\Lambda$ is unlikely to be successful,
because of the limited range of $Q^2$ available now and in the
foreseeable future. It is suggested
 that the independent determination of $\Delta G$ from charm
production is thus an essential element if $\Delta\bar u$,
$\Delta\bar d$, $\Delta s +\Delta\bar s$ are to be determined
accurately.
 Finally, motivated by the arguments that possibly $s(x)\not=\bar
s (x)$ and $\Delta s(x)\not=\Delta\bar s (x)$, we have
demonstrated how, in principle, one can learn about $s(x)-\bar s
(x)$ and $\Delta s(x)-\Delta\bar s (x)$.

The procedure we have advocated poses a real challenge to the
experimentalists, since it requires a control over the
 systematic errors involved in hadron detection efficiencies. The
price paid in the current practice of considering certain ratios
of cross sections in order to limit systematic errors, is
enormous, and vast amounts of interesting and theoretically
valuable information are thereby lost. We hope this paper will
encourage efforts to proceed further.

\section{Acknowledgements}

We thank M. Anselmino, H. Blok, M. Boglione, S. J. Brodsky,
 U. D'Alesio, D. de Florian, A. Kotzinjan, S. Kretzer, R. Sassot,
X. Song,
 D. Stamenov and C. Weiss for helpful discussions. We are
grateful to the UK Royal Society for a Collaborative Grant. E.C.
is grateful to the Theory Division of CERN for its hospitality
where this work was partly done and finished, her work was also
supported by the Bulgarian National Science Foundation. E.L. is
grateful
 to P.J. Mulders for the hospitality of the Department of Physics
and Astronomy, the Vrije Universitiet, Amsterdam, where part of
this work was carried out supported by the Foundation for
Fundamental Research on Matter
 (FOM) and the Dutch Organisation for Scientific Research (NWO).


\end{document}